\documentclass[aip,floatfix]{revtex4-1}

\usepackage{graphicx}
\usepackage{amsmath}

\newcommand{\eq}[1]{Eq.\ (\ref{#1})}

\begin{document}


\title{Evaluation of direct inversion of proton radiographs in the context of cylindrical implosions} 



\author{J. R. Davies}
\email[]{jdav@lle.rochester.edu}
\author{P. V. Heuer}
\affiliation{Laboratory for Laser Energetics, University of Rochester, Rochester, New York 14623, USA}



\date{\today}

\begin{abstract}
Direct inversion of deflectometry data, such as proton radiographs and shadowgraphs, is a well-posed problem with a unique solution for the transverse deflection of each particle or ray if their trajectories do not cross.
When trajectories cross, there exists an infinite set of solutions.
In proton radiography direct inversion determines the line-integrated transverse Lorentz force.
We have tested five publicly available direct inversion routines with a view to analyzing proton radiographs of cylindrical implosions on the OMEGA laser; four Monge-Amp\`{e}re solvers [github.com/flash-center/PRaLine, github.com/flash-center/PROBLEM, github.com/mfkasim1/invert-shadowgraphy/tree/fast-inverse, github.com/OxfordHED/proton-radiography-no-source], and a power-diagram method [github.com/mfkasim1/invert-shadowgraphy].
Test problems were generated using four field profiles, three cylindrical and one spherical, with varying field amplitudes in proton-tracing routines.  
Two Monge-Amp\`{e}re solvers did not run, the other two failed to reproduce radiographs when trajectories crossed, although for one field profile the solutions only diverged from the original at the boundary.
The power-diagram method was successful even when proton trajectories crossed, giving a solution that
minimized proton deflection, but failed for profiles that produced a single, sufficiently sharp peak.
For cases where trajectories do not cross, the Monge-Amp\`{e}re solvers have the advantage of being considerably faster than the power-diagram routine, up to $1000$ times in our tests. 
The test problems are provided in the supplementary information in pradformat [github.com/phyzicist/pradformat]. 
\end{abstract}


\maketitle 

\section{Introduction}
\label{sec-intro}
 
Over the last few decades, proton radiography has become a well-established diagnostic for electric and magnetic fields in laser-produced plasmas at multi-TW, multibeam laser facilities.
The protons are generated by either the interaction of laser pulses with a duration $\le10$ ps and relativistic intensities ($I\lambda^2 > 10^{18}$ W cm$^{-2}$ $\mu$m$^2$) with a foil or implosion of a D-$^3$He--filled sphere by multiple laser pulses, with a duration of the order of a nanosecond. 
A proton radiograph encodes the path-integrated transverse Lorentz force experienced by the protons, provided that proton scattering and energy loss due to collisions are negligible.
Collisions put an upper limit on the areal density for which proton radiography is a useful field diagnostic. 
While it is possible to use proton radiography to infer areal density under certain circumstances, it will not be considered here; we will consider the sub-field of proton radiography that could be termed ``proton deflectometry''.
Deflectometry can be extended to any charged particles and to shadowgraphy, where the transverse refractive index gradient determines the deflection.

The most-common approach to analyzing proton radiographs has been proton tracing in either hand-specified fields or fields from simulations, using a qualitative assessment of the agreement between the proton tracing and the data. 
Proton tracing determines if a proton radiograph is consistent with a given field profile, but it does not prove that the chosen field profile was the one present in the experiment.
An individual proton radiograph has the inherent limitations of being 2-D and not being able to distinguish between the electric and magnetic field components of the path-integrated transverse Lorentz force. 
Furthermore, when field gradients are sufficient to cause proton trajectories to cross one another a proton radiograph may be reproduced by a range of path-integrated transverse force profiles.

Recently, direct inversion of proton radiographs has been developed by a number of authors, using a number of different algorithms.\cite{Graziani,powerDiagram,Fast,PROBLEM,PRNS,machineLearning}   
If proton trajectories do not cross, direct inversion of a proton radiograph is a well-posed problem with a unique solution for the line-integrated transverse Lorentz force.
In general terms, this is a long-standing class of problems in physics and applied mathematics, first formulated in a paper published in 1781 by G. Monge,\cite{Monge} and direct inversion algorithms were developed long before laser-based proton radiography was developed.  
For example, the same algorithms have been applied to shadowgraphs of underdense laser plasmas to obtain the line-integrated refractive-index gradient.\cite{powerDiagram}
In most laser-plasma experiments of interest, proton trajectories do cross so the radiograph does not uniquely determine the line-integrated transverse Lorentz force.
However, direct inversion algorithms could still produce one possible solution, which would be valuable information.

As part of a project to develop laser-driven MagLIF (magnetized liner inertial fusion\cite{Slutz}) on OMEGA,\cite{miniMagLIF} often referred to as mini-MagLIF, we carried out proton radiography experiments to determine the compression of an applied axial magnetic field in preheated cylindrical implosions.
The intended objective was not met because protons deflected to one side by the compressed axial magnetic field could not be distinguished from the background.
However, the radiographs showed unexpected features that we eventually determined to be due to azimuthal magnetic field in the corona, which caused transverse proton deflection because the cylinder's axis was at an angle to the plane of the detector imposed by OMEGA's port geometry.
Here we describe our first steps in applying publicly available proton inversion routines to this data; learning to use them and understanding their limitations by applying them to test problems.
Since we are interested in analyzing cylindrical implosions we considered a number of test cases that differ from those published previously.
The analysis of our experimental data will be published elsewhere.         

The remainder of this article is organized as follows: Section \ref{sec-calculations} illustrates the basic principles of direct inversion of proton radiographs and features specific to tilted cylinders using the paraxial approximation and simplified 1-D electric and magnetic fields.
Section \ref{sec-profiles} describes our test field profiles, the physical motivations behind them, and how test radiographs were generated.
Section \ref{sec-codes} introduces the inversion algorithms and applies them to the test problems.
Section \ref{sec-conclusions} presents our conclusions.

\section{Proton radiography in 1-D cylindrical geometry}
\label{sec-calculations}

\begin{figure}[htb]
\includegraphics[width=0.45\textwidth]{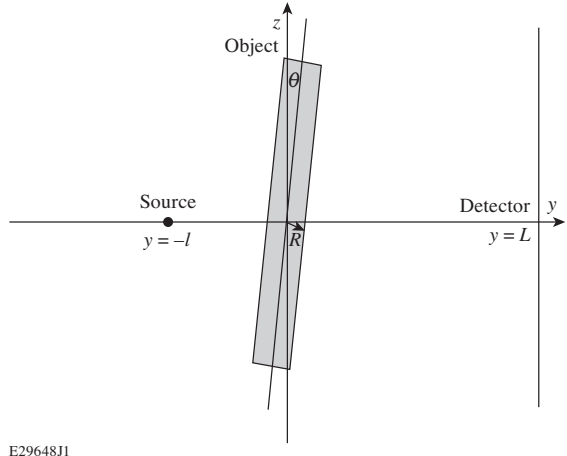}
\caption{\label{fig-setup} Schematic of the the 1-D cylindrical model considered in the text. The object is a cylinder of radius $R$ containing fields $E_r(r)$, $B_z(r)$, and $B_{\phi}(r)$ at a small angle $\theta$ to the $z$-axis, causing proton deflections in the $x$ direction, which is out of the page.}
\end{figure}

We are interested in determining electric and magnetic fields from proton radiographs of cylindrical implosions.
To gain some insight into the type of features we should expect specific to cylinders, in particular tilted cylinders, and to illustrate the basis of direct inversion of proton radiographs, here we consider the simplest possible model of paraxial proton propagation across an infinitely long cylinder where the only field components are $E_r(r)$, $B_z(r)$, and $B_{\phi}(r)$, where $r$ is the radial coordinate of the cylinder.  
In practice, any real system must have axial variations, leading to axial deflection of protons from line-integrated $E_z$ and $B_r$. 
However, near the middle of a cylinder of finite length with significantly smaller axial than radial gradients we can reasonably neglect axial deflection.  

The setup for this model is illustrated in Fig.\ \ref{fig-setup}.
The proton source is taken to be a point at a distance $l$ from the cylinder of radius $R$, which we place at the origin of our chosen coordinate system, and the detector is at a distance $L$ from the cylinder, giving a magnification
\begin{equation}
\label{eq-M}
M=\frac{L+l}{l}. 
\end{equation} 
We take the $x$ axis to be the direction resolved by the radiograph in the object plane, perpendicular to the axis of the cylinder, the $y$ axis to be the direction of probing, and the $z$ axis to be the axis of the cylinder as seen from the detector. 
Since the cylinder in our experiments was tilted, we will consider a cylinder at an angle $\theta$ to the $z$ axis in the $y$-$z$ plane, small enough that the change in magnification along the region of interest is negligible.

Practically, we require $l \gg R$ so that the laser-based proton source and the laser-plasma experiment do not interfere with one another; therefore the trajectories of protons reaching the cylinder will always be approximately parallel to the $y$ axis.  
If the deflection of the protons in $x$ within the cylinder is sufficiently small, it can be determined approximately from the integral in $y$ of the $x$ component of the force; in other words, we can use the paraxial approximation
\begin{equation}
\label{eq-Dp_p}
\frac{\Delta v_x}{v} \approx \frac{1}{2 \rm E}\int F_x dy = \frac{\mathcal{F}_x}{2\rm E} \ll 1,
\end{equation}  
where $v$ is velocity, E is energy and we introduce $\mathcal{F}$ for line-integrated force.
We distinguish ``path-integrated'' as the force integrated along the actual, self-consistent proton trajectory and ``line-integrated'' as the force integrated parallel to the probing axis.
In terms of the electric and magnetic fields considered in our model, we have
\begin{equation}
\label{eq-Fr}
\mathcal{F}_x = 2q\int_x^R \left( \frac{xE_r}{r\cos\theta} + \frac{xvB_{\phi}\tan\theta}{r} + vB_z  \right) \frac{rdr}{\sqrt{r^2-x^2}},
\end{equation}
where $q$ is proton charge. 
Equation (\ref{eq-Fr}) is an Abel transform.
A $1/\cos\theta$ factor arises from the increase in path length across the tilted cylinder, and the apparent axial magnetic field seen by the protons is $B_z\cos\theta + B_{\phi}\sin\theta$.
Equation (\ref{eq-Fr}) illustrates the important point that for a tilted cylinder azimuthal magnetic field will cause a transverse proton deflection in the same manner as a radial electric field.

We also require $L \gg R$; therefore we can determine the approximate position a proton lands on the detector $x_{\rm d}$ by assuming that all of the deflection given by Eq.\ (\ref{eq-Dp_p}) occurs at $y=0$, giving
\begin{equation}
\label{eq-xd}
x_{\rm d} \approx  x+\frac{L\mathcal{F}_x}{2M{\rm E}},
\end{equation}
where distance is given in object plane equivalent units (physical distance on the detector divided by magnification $M$), so
due to the paraxial approximation the coordinate $x$ at which a proton crosses the cylinder and the position $x$ at which an undeflected proton lands on the detector are identical. 
Equation (\ref{eq-xd}) gives the proton deflection at the detector, therefore, we can use it to transform from the proton intensity for undeflected protons $I_0$ to the measured proton intensity $I$
\begin{equation}
\label{eq-I_I0}
\frac{I}{I_0} \approx \frac{1}{\left| 1+\frac{L}{2{\rm E}M}\frac{d\mathcal{F}_x}{dx}  \right|}, \quad \frac{L}{2{\rm E}M}\frac{d\mathcal{F}_x}{dx} \ll 1,
\end{equation}  
provided that Eq.\ (\ref{eq-xd}) is a differentiable, single-valued function of $x$.

Equation (\ref{eq-I_I0}) provides four important physical insights: First, for sufficiently small field gradients, the modulation in the proton intensity is directly related to the line-integrated transverse Lorentz force on the protons, demonstrating that direct inversion of a proton radiograph to obtain the line-integrated force is possible.
Second, direct inversion will require the proton intensity in the absence of forces $I_0$.
Typically $I_0$ will not be uniform.
For our 1-D cylindrical test problems with a source that has a uniform angular distribution
\begin{equation}
\label{eq-I0cyl}
I_0 \propto \frac{1}{1+x^2/l^2},
\end{equation}
where $x$ is in object plane equivalent units. 
For our test problems we deliberately chose $l$ large enough that $I_0$ is uniform to better than 1\% over the region of interest.
For a point source with a uniform distribution in solid angle, the typical case for actual proton radiography, we have
\begin{equation}
\label{eq-I0}
I_0 \propto \frac{1}{(1+x^2/l^2+z^2/l^2)^{3/2}}.
\end{equation}
Laser-foil proton sources have been found to have narrow, non-uniform distributions in solid angle that can vary from shot to shot,\cite{PRNS} which is a major issue for quantitative analysis of proton radiographs. 
Our experiments used a D-$^3$He fusion source, which has been found to have a reproducible, uniform distribution in solid angle.
Third, if Eq.\ (\ref{eq-I_I0}) diverges then proton trajectories cross and there no longer exists a unique solution for the line-integrated transverse Lorentz force for a given proton intensity modulation.
Points where the intensity distribution theoretically diverge are known as caustics. 
Finally, a natural dimensionless measure of the force $F$ on the proton is
\begin{equation}
\label{eq-mu}
\mu=\frac{LF}{M{\rm E}}.
\end{equation}
We have adopted the symbol $\mu$ because it has been adopted as a dimensionless measure of proton deflection in a number of other papers on proton radiography, as discussed by Bott {\it et al.}\cite{PROBLEM}
The definitions of $\mu$ used differ however and are equivalent only for $\mu \ll 1$. 
Other definitions rely on the deflection of a proton,\cite{PROBLEM} but when $\mu$ is not small, the deflection can only be accurately determined by proton tracing in the fields. 
Our definition allows a direct determination of $\mu$ from specified fields, which is the problem we are considering here.
By adopting dimensionless distances using an appropriate object size, which for our cylindrical case we will write as $R$, we can produce dimensionless radiographs for a given radial Lorentz force profile characterized simply by $\mu_{\max}$ applicable to any deflectometry setup with $l \gg R$ and $L \gg R$. 

\section{Test profiles}
\label{sec-profiles}

We generated test radiographs for four radial force profiles with maximum, absolute values of $\mu_{\max}$ [\eq{eq-mu}] varying from $1/8$ up to $4$; three cylindrical, giving 1-D radiographs, and one spherical, giving 2-D radiographs.
Here we give the radial force profiles, the line-integrated transverse force profiles, the results of using the paraxial approximation to calculate the intensity modulations [\eq{eq-I_I0}], sample radiographs, and discuss the motivations for choosing these profiles.

The most widely used test case is a spherical Gaussian potential, therefore we considered a spherical and a cylindrical Gaussian potential, which both have dimensionless radial force profiles
\begin{equation}
\label{eq-mu_test}
\mu_r = \mu_{\max} \sqrt{2{\rm e}} \frac{r}{R}\exp\left( -\frac{r^2}{R^2} \right).
\end{equation}
Since the spherical Gaussian potential has been extensively considered in numerous publications, we will not consider it in any detail here.
Equation (\ref{eq-mu_test}) is a physically reasonable profile for the radial electric field and azimuthal magnetic field in a tilted cylinder, where
\begin{equation}
\label{eq-muEB}
\mu_r = \frac{qE_rL}{M{\rm E}\cos\theta} + \frac{2qB_{\phi}L\tan\theta}{Mp},
\end{equation}
where $p$ is proton momentum.
The line-integrated transverse Lorentz force for the cylindrical Gaussian potential is 
\begin{equation}
\mathcal{F}_x = \mathcal{F}_{\max} \sqrt{2{\rm e}} \frac{x}{R}\exp\left( -\frac{x^2}{R^2} \right).
\end{equation}
Test radiographs for the cylindrical Gaussian were obtained by numerical integration of the equation of motion using a fourth-order Runge-Kutta scheme with a direct calculation of the field on the protons and randomly distributed protons, with a bin width of $0.05R$ and a mean of at least 1000 particles per bin.
Test radiographs for the spherical Gaussian were obtained using PlasmaPy,\cite{PlasmaPy} which uses fields calculated on a 3-D Cartesian grid interpolated to the protons using first-order weighting, randomly distributed protons, and time-centered numerical integration of the equation of motion, with a bin width of $0.052R$ and a mean of 10 particles per bin.

\begin{figure}[htb]
\includegraphics[width=0.45\textwidth]{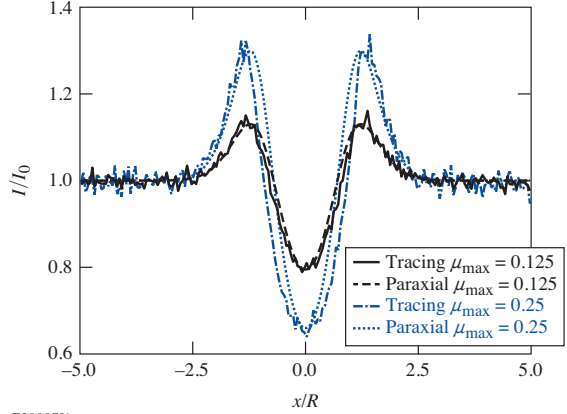}
\caption{\label{fig-paraxialComp} Sample proton radiographs for a cylindrical Gaussian potential obtained by numerical integration of randomly selected proton trajectories using a mean of 1000 particles per bin with a bin width of $0.05R$ compared to the analytical, paraxial prediction of \eq{eq-Gauss}.}
\end{figure}

Applying Eq.\ (\ref{eq-I_I0}) to the cylindrical Gaussian, with the upper limit of the integration tending to infinity, gives
\begin{equation}
\label{eq-Gauss}
\frac{I}{I_0} = \frac{1}{\left| 1+ \mu_{\max} \sqrt{\frac{\pi \rm e}{2}} \exp\left( -x^2/R^2\right) \left( 1-2x^2/R^2\right) \right|}.
\end{equation}
Although technically the radial extent of the fields is infinite, which does not satisfy the paraxial approximation, in practice the vast majority of the deflection occurs within a few e foldings. 
As can be seen in Fig.\ \ref{fig-paraxialComp}, at $\mu_{\max}=1/8$ Eq.\ (\ref{eq-Gauss}) is adequate, whereas at $\mu_{\max}=1/4$ there are already significant discrepancies in the width of the depression and the position of the peaks. 

\begin{figure}[htb]
\includegraphics[width=0.45\textwidth]{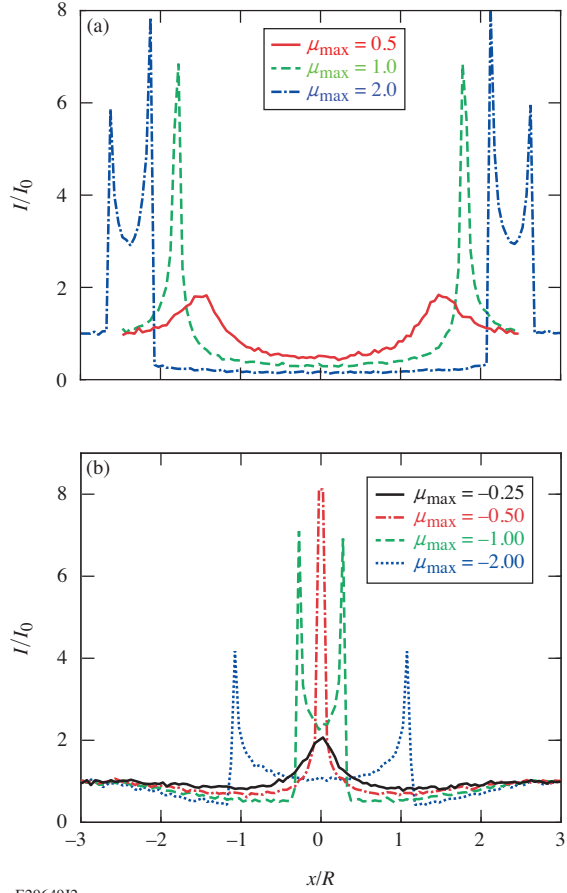}
\caption{\label{fig-caustics} Sample proton radiographs for a cylindrical Gaussian potential obtained by numerical integration of randomly selected proton trajectories using a mean of 1000 particles per bin with a bin width of $0.05R$.}
\end{figure}

\begin{figure}[htb]
\includegraphics[width=0.45\textwidth]{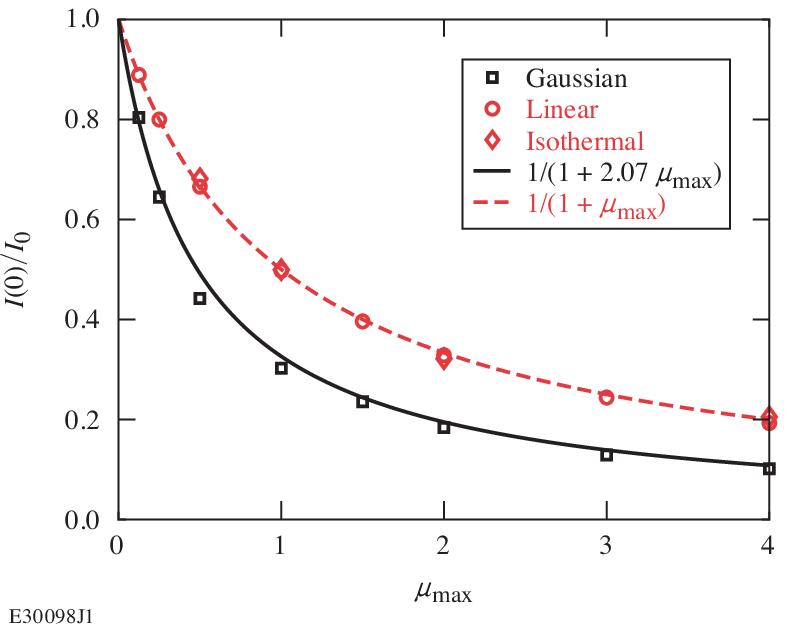}
\caption{\label{fig-I0} Proton intensity depression on axis from proton tracing for the cylindrical Gaussian, linear, and isothermal $\ln\Lambda = 10$ profiles, and the predictions for the cylindrical Gaussian from \eq{eq-Gauss} and the linear profile from \eq{eq-E_propto_r}.}
\end{figure}

Equation (\ref{eq-Gauss}) diverges when
\begin{equation}
\label{eq-mu_cross}
\mu_{\max} \ge \frac{\rm e}{\sqrt{2\pi}} \approx 1.08, \ \mu_{\max} \le -\sqrt{\frac{2}{\pi \rm e}} \approx -0.484.
\end{equation}
Note that both negative and positive values of $\mu_{\max}$ will occur for $B_{\phi}$, while we expect $E_r$ to be positive in a compressed cylindrical plasma, but not for all possible experiments.
Clearly, negative values, which correspond to focusing fields, cause the paraxial approximation to break down and trajectories to cross at lower field amplitudes.
As can be seen from Fig.\ \ref{fig-caustics}, Eq.\ (\ref{eq-mu_cross}) accurately predicts the onset of caustics.
Even when $\mu_{\max} \ge 1.08$, Eq.\ (\ref{eq-Gauss}) accurately predicts the intensity depression on axis, as shown in Fig.\ \ref{fig-I0}, because the paraxial approximation remains valid close enough to the axis.
 
The first cylindrical profile we considered was Murakami and Basko's \cite{MB} self-similar solution for the radial electric field in an isothermal electron expansion with cold ions.
It is not possible to find an explicit solution for the radial electric field from the cylindrical Murakami-Basko equation, but we found an adequate approximation:
\begin{eqnarray}
\label{eq-Eiso}
E_r & \approx & \frac{r}{R}E_{\max},\phantom{\exp[\ln\Lambda(1-r^2/R^2)]} \qquad r \le R, \nonumber \\
     & \approx & \frac{R}{r}\exp[\ln\Lambda(1-r^2/R^2)]E_{\max}, \qquad r > R, 
\end{eqnarray}
where $R$ is the position of the ion front, and $\Lambda$ is a dimensionless parameter that characterizes the electron sheath ahead of the ion front (not the $\Lambda$ of transport theory).
More details are given in the appendix.

\begin{figure}[htb]
\includegraphics[width=0.45\textwidth]{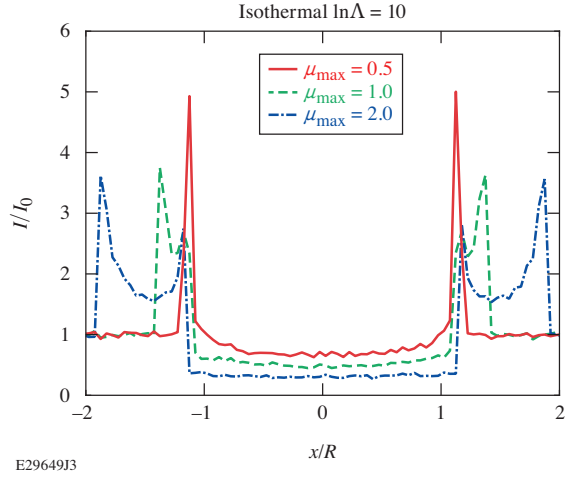}
\caption{\label{fig-isothermal} Sample proton radiographs for an isothermal cylindrical expansion obtained by numerical integration of randomly selected proton trajectories using a mean of 1000 particles per bin with a bin width of $0.05R$.}
\end{figure}

Sample proton radiographs for the isothermal force profile are given in Fig. \ref{fig-isothermal}, which were obtained in the same manner as the cylindrical Gaussian.
A key feature of this profile is that the discontinuity in the field at the ion front always leads to proton trajectories crossing, causing a sharp spike, or caustic, in the protons just beyond the ion front. 
Therefore, we should expect proton trajectories to cross in radiographs of plasma expanding into vacuum. 
For values of $\mu_{\max} > 1$ a second spike appears at a larger radius that increases with $\mu_{\max}$.
For $\mu_{\max} > 1.08$ both the cylindrical Gaussian potential and the isothermal expansion give radiographs with broad, flat intensity depressions and double caustics, giving an example of the degeneracy in solutions when trajectories cross. 
In this case, it would make it difficult to be certain whether a caustic is indicative of an ion front or not.

\begin{figure}[htb]
\includegraphics[width=0.45\textwidth]{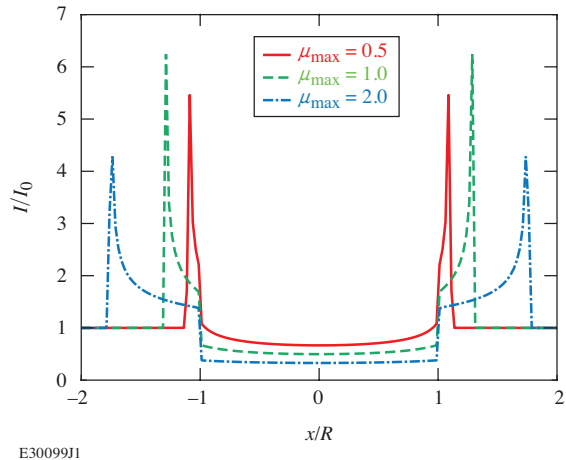}
\caption{\label{fig-linear} Sample proton radiographs for the linear profile ($\mu_r \propto r$ at $r \le R$) obtained from the analytic solutions for uniformly distributed proton trajectories using a mean of 1000 particles per bin with a bin width of $0.025R$.}
\end{figure}

The isothermal profile has a number of disadvantages as a test case: the line-integrated force cannot be determined analytically, Eq.\ (\ref{eq-I_I0}) cannot be used to obtain the proton modulation in the paraxial limit, and generating the proton radiographs requires some care in dealing with crossing of the ion front and small step sizes in the sheath.
Therefore, we decided to neglect the rapidly decaying electric field in the sheath, giving what we will call a linear profile because it has $\mu_r \propto r$ for $r \le R$, giving a line-integrated transverse Lorentz force of 
\begin{equation}
\mathcal{F}_x = \mathcal{F}_{\max} 2 \frac{x}{R} \sqrt{\max(1-x^2/R^2,0)}.
\end{equation}
Test radiographs for the linear profile were obtained using the analytic solution for proton trajectories and uniformly distributed protons, allowing the rapid generation of almost noise free profiles without grid imprinting. 
A bin width of $0.025R$ was used with a mean of at least 1000 particles per bin. Sample test profiles are shown in Fig.\ \ref{fig-linear}.
For a linear profile Eq.\ (\ref{eq-I_I0}) gives
\begin{eqnarray}
\label{eq-E_propto_r}
\frac{I}{I_0} & = &  \frac{1}{\left| 1+ \mu_{\max} (1-2x^2/R^2)/\sqrt{1-x^2/R^2} \right|}, \qquad x \le R, \nonumber \\
 & = & 1, \qquad x>R.
\end{eqnarray}
Equation (\ref{eq-E_propto_r}) always diverges because of the discontinuity in the gradient of the force at $R$, therefore it cannot be used to determine the proton intensity modulation however small $\mu_{\max}$, but it does give an accurate measure of the intensity depression on axis for $\mu_{\max} > 0$, which is also adequate for the isothermal profile, as seen in Fig.\ \ref{fig-I0}. 

The objective of the experiments that motivated this study was to measure proton deflection by a compressed axial magnetic field.
The compressed axial magnetic field is expected to be discontinuous at the shell-gas interface and sharply peaked near the axis.\cite{Us} 
The simplest possible approximation is a top-hat axial magnetic field profile
\begin{eqnarray}
B_z & = & B_{\max}, \qquad r \le R, \nonumber \\
\label{eq-B}
      & = & 0, \qquad r > R.
\end{eqnarray}
The line-integrated transverse Lorentz force is given by
\begin{equation}
\mathcal{F}_x = \mathcal{F}_{\max} \sqrt{\max(1-x^2/R^2,0)}.
\end{equation}
From Eq.\ (\ref{eq-I_I0}) we obtain
\begin{eqnarray}
\label{eq-I_I0tophat}
\frac{I}{I_0} & = & \frac{1}{\left|1-\mu_{\max}(x/R)/\sqrt{1-x^2/R^2}\right|}, \qquad x \le R, \nonumber \\
                    & = & 1, \qquad x>R.
\end{eqnarray}
which always diverges because of the discontinuity in the force. 
However, Eq.\ (\ref{eq-I_I0tophat}) does show that for $\mu_{\max} \ll 1$ the axial magnetic field could be determined from the slope of the proton intensity modulation near the axis. 

\begin{figure}[htb]
\includegraphics[width=0.45\textwidth]{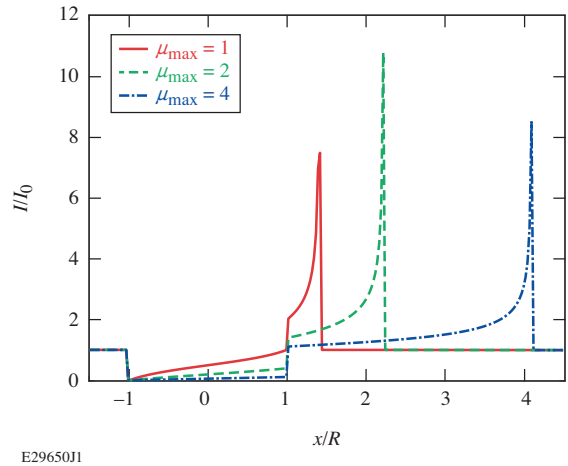}
\caption{\label{fig-Bz} Sample proton radiographs for a top-hat radial force profile [Eq.\ (\ref{eq-B})] obtained using an analytic solution for the proton trajectories with a mean of 1000 uniformly distributed protons per bin and a bin width of $0.025R$.}
\end{figure}

Test radiographs for the top-hat profile were obtained using the analytic solution for proton trajectories and uniformly distributed protons, with a bin width of $0.025R$ and a mean of at least 1000 particles per bin.
Sample radiographs are shown in Fig.\ \ref{fig-Bz}. 
For $\mu_{\max}> 1$, the regime relevant to our experiments, practically all of the protons are deflected out of the core into a peak with a long tail at a distance that increases approximately linearly with $\mu_{\max}$.
An adequate calculation of the position of the peak for $\mu_{\max} \gg 1$ is given by the deflection of a proton traveling along the axis,
\begin{equation}
x_{\rm p} \approx  \frac{\mu_{\max}R}{1-R^2/r_{\rm L}^2},
\end{equation}  
where $r_{\rm L}$ is Larmor radius, given by $p/qB_z$.
Therefore, the peak radially integrated axial magnetic field $B_zR$ can be inferred by determining the position of a single peak that appears on only one side of the radiograph, determined by the direction of the axial magnetic field.
The value of $B_zR$ is an important metric for magneto-inertial fusion,\cite{Slutz} so it is important to obtain a direct measurement. 
Unfortunately, in our experiments this peak could not be distinguished from the background.
However, the top-hat profile provides a tough test for proton inversion routines. 
The asymmetric intensity modulation is quite different to any of our other profiles and any profile used in previously published tests.
  
\section{Evaluation of direct inversion algorithms}
\label{sec-codes}

We found five direct inversion routines publicly available on GitHub.\cite{Graziani,powerDiagram,Fast,PROBLEM,PRNS}
All of them output a deflection potential $\phi$, defined by
\begin{equation}
\mathcal{F} = -\nabla \phi.
\end{equation}
PRaLIne\cite{Graziani}, PROBLEM\cite{PROBLEM}, PRNS\cite{PRNS} and fast\_invert\_shadowgraphy\cite{Fast} solve the Monge-Amp\`{e}re equation.\cite{Monge}
PROBLEM, PRNS and fast\_invert\_shadowgraphy implement the Sulman, Williams and Russell algorithm,\cite{Sulman} which changes the problem to finding the steady-state solution to a diffusion-like equation. 
The most general form of the Monge-Amp\`{e}re equation is given by the determinant of the Jacobian of $-\nabla \phi$ equal to a function of position and $\nabla \phi$.
We can illustrate the origin of the Monge-Amp\`{e}re equation by using Eq.\ (\ref{eq-I_I0}) to obtain
\begin{equation}
\frac{d^2\phi}{dx^2} = \frac{2{\rm E}M}{L} \left( 1-\frac{I_0}{I} \right)
\end{equation}
which is a Poisson equation, a special case of the Monge-Amp\`{e}re equation where the source is independent of $\nabla \phi$.
Poisson solvers have been applied to shadowgraphy,\cite{powerDiagram} where it was found that they only work for very small modulations; Fig.\ \ref{fig-paraxialComp} shows that the paraxial approximation is inadequate for a Gaussian potential at $\mu_{\max}=0.25$.
In order to obtain an equation that is adequate for larger modulations it is necessary to go beyond the paraxial approximation, which leads to the source term depending on $\nabla \phi$, giving the Monge-Amp\`{e}re equation.
The full derivation is too lengthy to repeat here, but can be found in a number of publications.\cite{Graziani,PROBLEM,Monge}

PRNS differs from the other routines in that it can determine a probability distribution for the source intensity starting from a given prior, assuming that there exists a unique solution for the line-integrated transverse Lorentz force.
All of our test problems have uniform source intensities, so we did not test the no source capability of PRNS.
Therefore, after verifying that PRNS gave similar results to fast\_invert\_shadowgraphy for some test cases, as expected since it uses the same algorithm implemented by the same author, we did not use it further.
Some users may prefer PRNS since it is written in Python rather than Matlab.
The Python codes for PraLIne and PROBLEM we obtained from GitHub did not run, so we will not consider them here. 
Therefore, the only Monge-Amp\`{e}re routine we will show results for is fast\_invert\_shadowgraphy, so we will refer to it simply as Monge-Amp\`{e}re.

The routine invert\_shadowgraphy,\cite{powerDiagram} uses established algorithms from computational geometry. 
The routine starts by constructing a Voronoi diagram of the source intensity from randomly selected sites, using a rejection algorithm based on the source if it is non-uniform.
In a Voronoi diagram, a cell is the region that contains the points closer to the given site than any other site. 
The sites are then iteratively replaced with the centers-of-mass of their cells to approach equal flux in each cell (considering flux as mass).
The selected sites are then used to construct a power-diagram of the data, a weighted Voronoi diagram, starting from no weighting and then iteratively adjusting the weights using a minimization algorithm to approach equal flux in each cell.
Finally, the displacement of the centers-of-mass of the weighted cells from the original source sites are used to determine the line-integrated forces, which in turn are used to calculate the deflection potential. 
We will refer to this routine as the power-diagram routine.

All three of the routines we used output the solution as the same dimensionless deflection potential $\phi^{\prime}$.
The line-integrated force in the convenient units of MeV, which is also the line-integrated electric field in MV, at bin $i$ is given by 
\begin{equation}
\mathcal{F}_{i} = 2 {\rm E_{MeV}} \frac{wM}{L} (\phi_{i-1}^{\prime}-\phi_{i}^{\prime}) \mbox{ MeV},
\end{equation}
where E$_{\rm MeV}$ is the proton energy in MeV and $w$ is bin width in the object plane ($wM$ is bin width on the detector), in the same units as the object to detector distance $L$. 
The line-integrated magnetic field, assuming that there is only a magnetic field, in the convenient units of T mm is given by 
\begin{equation}
\mathcal{B}_{i} = 145 \sqrt{\rm E_{MeV}} \frac{wM}{L} (\phi_{i-1}^{\prime}-\phi_{i}^{\prime}) \mbox{ T mm}.
\end{equation}

None of the inversion routines are equipped to deal with 1-D problems and require a 2-D input.
We replicated our 1-D arrays an odd number of times until the deflection potential along the center appeared to converge.
In effect, we produced a 2-D radiograph for an axially uniform cylinder of finite length.
We then obtained a 1-D reconstruction of the radiograph by replicating the deflection potential along the center three times and putting it into the forward (reconstruction) routines. 
If the 1-D reconstruction appeared to match the original we accepted the solution.
We found that the Monge-Amp\`{e}re forward routine was less prone to noise than that of the power-diagram routine, so we used it in all cases.

The replication approach worked for the power-diagram routine with as few as 5 replications for the lower values of $\mu_{\max}$.
Taking the mean of the 2-D potential away from the boundary also gave an accurate 1-D solution, in some cases a more accurate one.
The results shown here were all obtained using 11 replications and taking the potential along the center.
It is interesting to note that the 2-D reconstructions were not accurate, always showing axial oscillations due to a noisy axial force.
A buffer of at least a few rows was always required because the power-diagram routine distorts the potential at the corners; at the end of the minimization step the vertices near the corners are moved back to the corners in order to obtain a potential over the whole grid.
The greater the deflection of the points near the corners the further this distortion spreads.

The replication approach did not work for the Monge-Amp\`{e}re routine because the boundary conditions assume that proton modulation goes to zero at the edges of the radiograph and errors at the boundary propagate over the entire grid since it is solving a diffusion-like equation.
We padded the replicated array with rows of uniform intensity, in effect adding a vacuum region at the ends of the finite cylinder, and this worked.
We found two rows of padding at each end to be optimal in terms of speed and accuracy of solution.
An adequate 1-D solution was only obtained when the number of replications was roughly equal to the number of points in the original array.
The results shown here were all obtained using an odd number of replications one greater than the number bins in the original 1-D profile, which was always even.
An adequate 1-D solution was never obtained for the top-hat profile and for the cylindrical Gaussian when caustics were present.
We tried producing a 2-D array using an envelope function
\begin{equation}
\frac{I}{I_0}(x,z) = 1 + f(z)\left(\frac{I}{I_0}(x)-1\right),
\end{equation}
with $f(z)$ given by $\exp(-z^{2n})$ with $f(\pm z_{\max})<10^{-3}$ and $\cos^2(z^2)$ with $z_{\max}=\pi/2$, so that the original radiograph was reproduced at the center ($z=0$) with zero derivative in $z$ and the modulation and the derivative tended to zero or were zero at $\pm z_{\max}$, but this did not change the results.
It is interesting to note that the 2-D reconstructions were always accurate away from the ends of the cylinder. 
Unlike the power-diagram routine, the axial force was smooth and essential to an accurate reconstruction, despite the radiograph being generated with zero axial force.
We believe the lack of convergence to a 1-D solution is due to the inability of the Monge-Amp\`{e}re routine to invert these profiles caustics.
This was confirmed by the failure of the Monge-Amp\`{e}re routine to invert the 2-D spherical Gaussian test problems with caustics.

For cylindrical and spherical Gaussian potentials with $-0.484 <\mu_{\max} < 1.08$ (no caustics) all of the inversion routines accurately reproduced the original radiographs and the original line-integrated forces, as expected.

\begin{figure}[htb]
\includegraphics[width=0.45\textwidth]{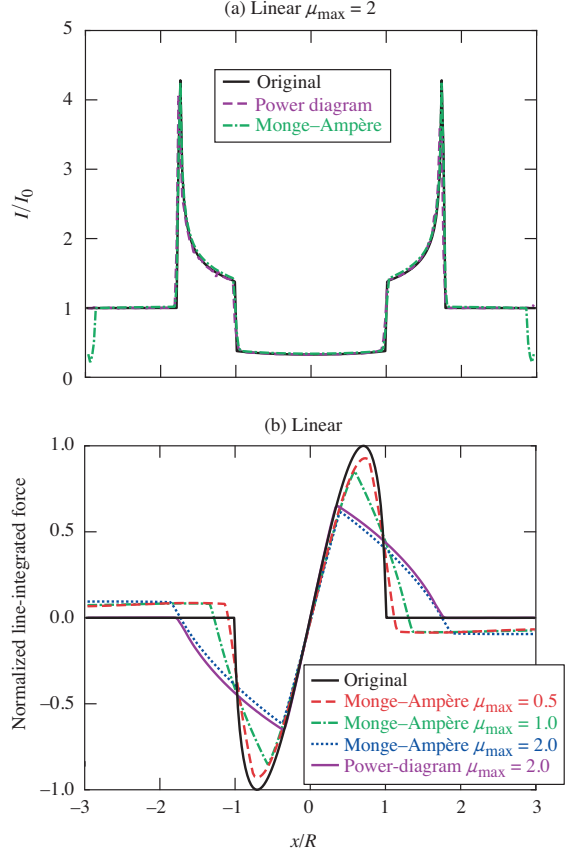}
\caption{\label{fig-MAlinear} Tests with the linear force profile, (a) reconstructions of the test radiograph for $\mu_{\max} = 2$ and (b) line-integrated forces normalised so that the maximum of the original is $1$ for a range of values of $\mu_{\max}$.}
\end{figure}

While the Monge-Amp\`{e}re routine failed to converge to a 1-D solution for the cylindrical Gaussian tests with caustics and the top-hat profile at all values of $\mu_{\max}$, it did give an adequate reconstruction for the linear profile.
An example of such a reconstruction is shown in Fig.\ \ref{fig-MAlinear}(a) for $\mu_{\max}=2$, and the line-integrated forces for a range of values of $\mu_{\max}$ up to $2$ are shown in Fig.\ \ref{fig-MAlinear}(b).
Results from the power-diagram for $\mu_{\max}=2$ are included for comparison.
For all values of $\mu_{\max}$, the Monge-Amp\`{e}re routine gave dips in the proton intensity at the edge that are not present in the original.
However, if boundary conditions of continuous force rather than zero force were applied it should match the original.
The reconstruction would tend to the original if the boundaries were moved to infinity.
In this case, the uniform fields at the edge seen in Fig.\ \ref{fig-MAlinear}(b) would extend to infinity, which is clearly not a physically acceptable solution for a finite, net neutral system.
Nonetheless, the Monge-Amp\`{e}re routine can be considered to have found a solution for the linear profile, which differs from the original profile because there does not exist a unique solution.
Figure \ref{fig-MAlinear}(b) therefore shows three possible solutions for the linear profile with $\mu_{\max}=2$, the original, the Monge-Amp\`{e}re solution and the power-diagram solution, providing an excellent illustration of the degeneracy in solutions once trajectories cross.

\begin{figure}[htb]
\includegraphics[width=0.45\textwidth]{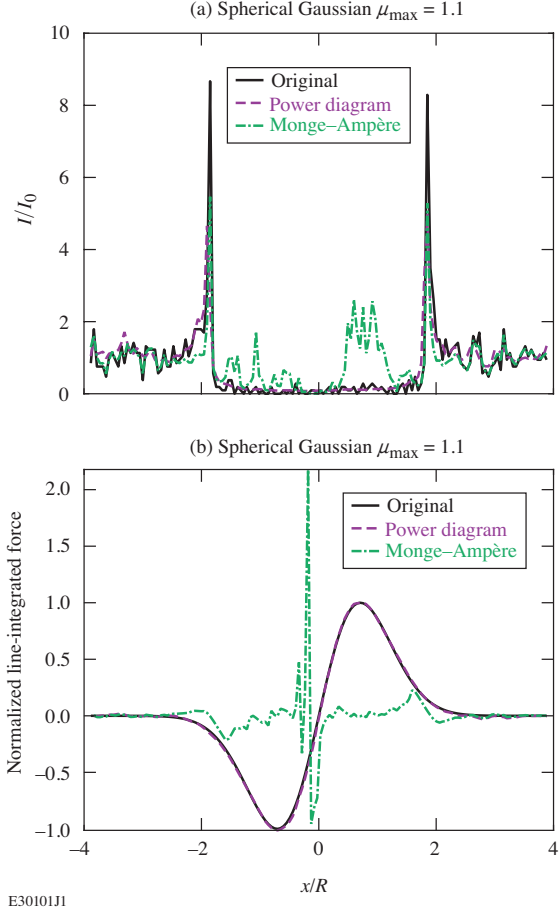}
\caption{\label{fig-failure} Line-outs through the center of results for the spherical Gaussian potential with $\mu_{\max}=1.1$ for (a) the proton radiographs, and (b) the line-integrated fields normalized so that the maximum of the original is $1$.}
\end{figure}

\begin{figure}[htb]
\includegraphics[width=0.9\textwidth]{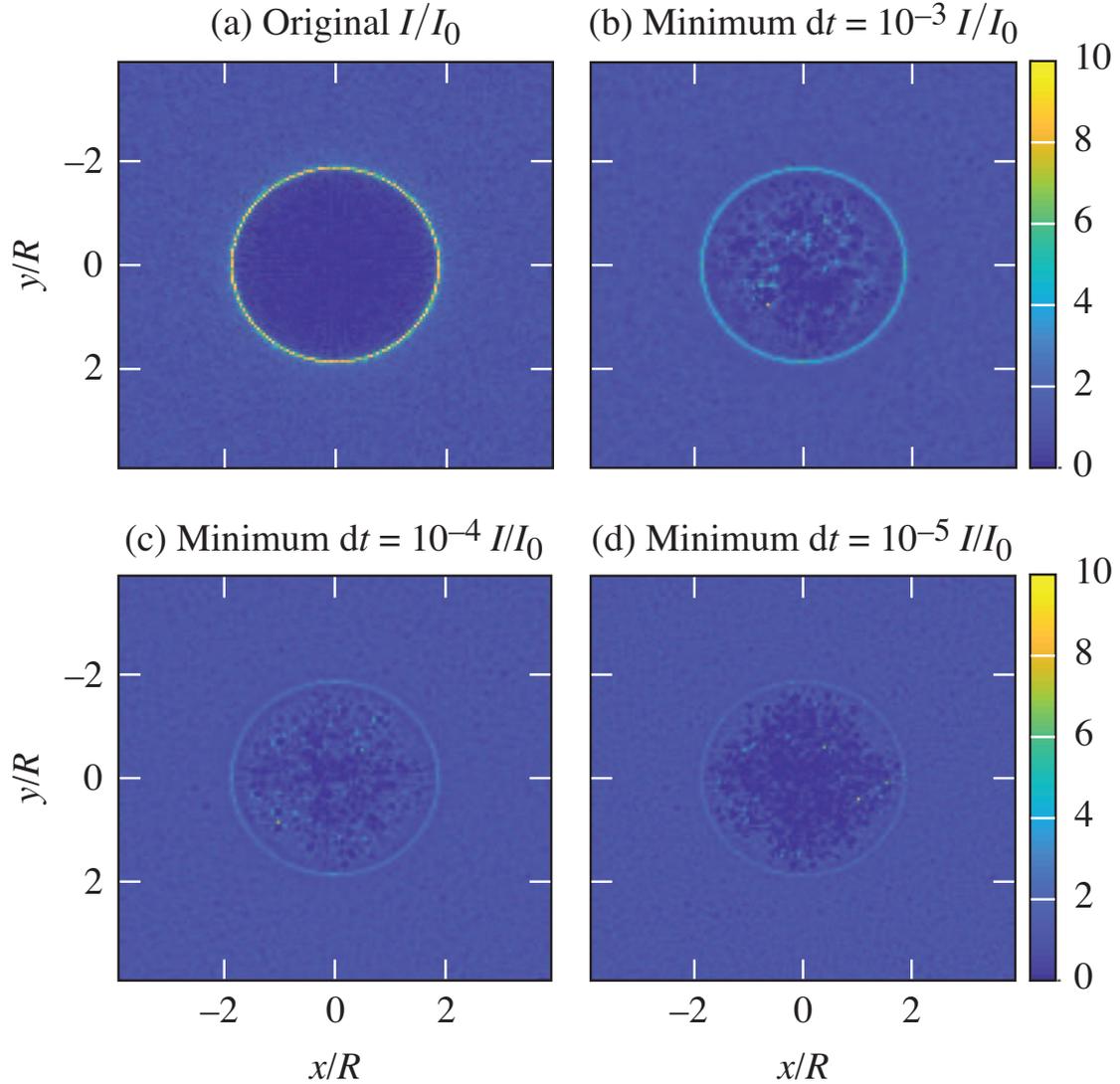}
\caption{\label{fig-step}Study of the effect of the minimum time step in the Monge-Amp\`{e}re routine on the reconstruction of the spherical Gaussian with $\mu_{\max}=1.1$, (a) original test radiograph, (b) reconstruction with default parameters, (c) with minimum time step reduced by a factor of $10$, and (d) with minimum time step reduced by a factor of $100$}
\end{figure}

An example of the Monge-Amp\`{e}re routine failing to produce an accurate solution for a spherical Gaussian potential once caustics appear is shown in Fig.\ \ref{fig-failure}.
We tried reducing the minimum time step in the Monge-Amp\`{e}re routine, as shown in Fig.\ \ref{fig-step}, but never obtained an accurate reproduction of the original radiograph.
We found the minimum time step to have the most significant effect on the solution out of the numerical parameters in the main\_inverse routine, which are relative tolerance, minimum step, alpha, and interpolation and extrapolation methods, which can be nearest or linear.
We found that decreasing relative tolerance by a factor of up to $100$ in our step size scan made no noticeable improvement. 
We varied alpha by a factor of $2$ either way on one case and it made no significant differences. 
We did not change interpolation and extrapolation methods from ``nearest'' because the comments in the code state that this is more robust.

\begin{figure}[htb]
\includegraphics[width=0.9\textwidth]{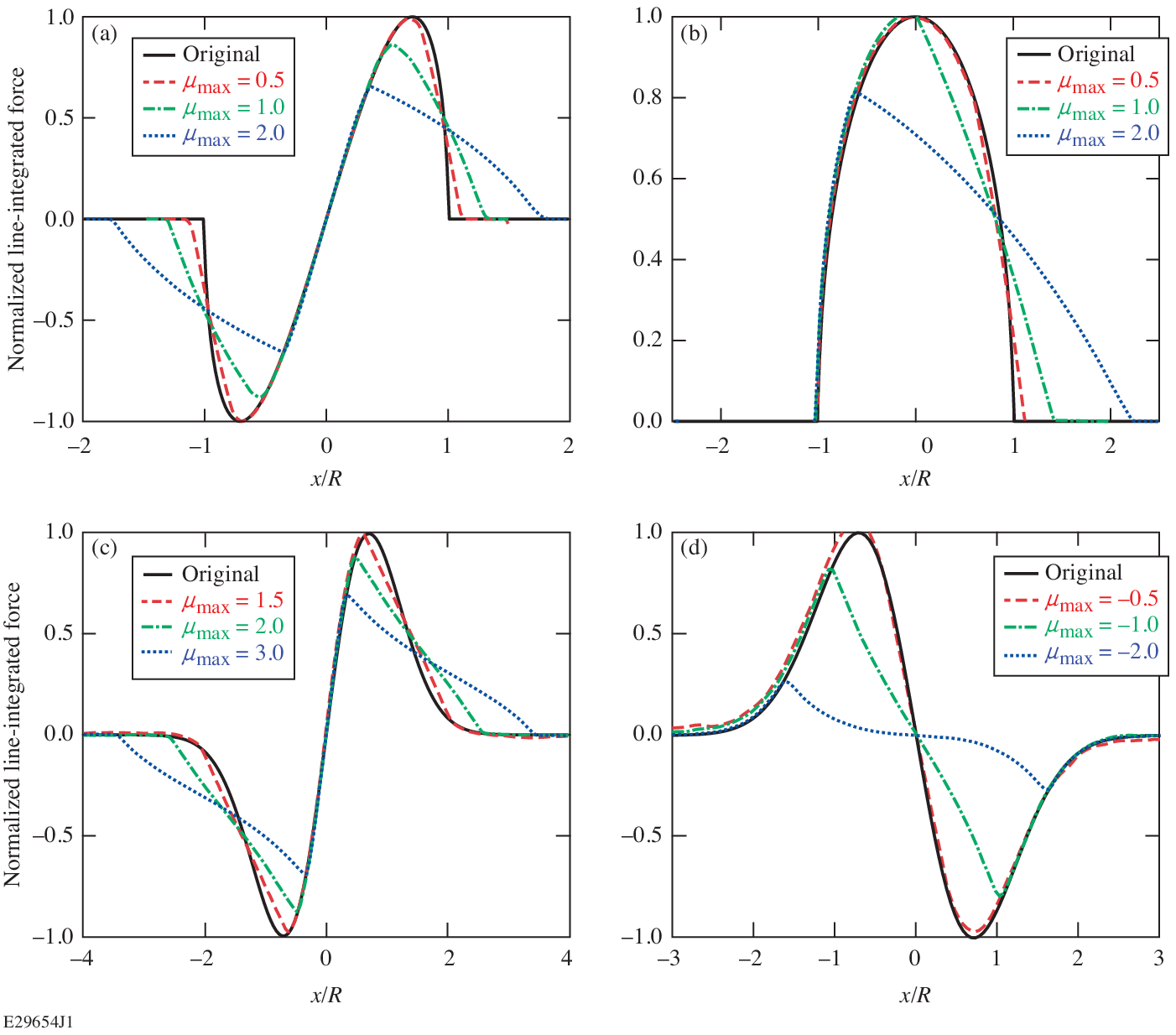}
\caption{\label{fig-PowerDiagram} Line-integrated fields from the power-diagram inversion routine, normalized so that the peak of the original value is $1$, for (a) the linear force profile, (b) the top-hat force profile, (c) the cylindrical Gaussian potential for $\mu_{\max} > 0$ and (d) the cylindrical Gaussian potential for $\mu_{\max} < 0$ at select values of $\mu_{\max}$. The bin widths were $0.025R$ for the linear and top-hat profiles, and $0.05R$ for the Gaussian potential. In all cases the original proton radiograph was adequately reproduced by the power-diagram routine.}
\end{figure}

The power-diagram routine successfully inverted all of the cylindrical test profiles, in that it provided an adequate match to all key features in the original radiographs, within the noise level.
Examples of the line-integrated forces obtained for all of  the cylindrical profiles are given in Fig.\ \ref{fig-PowerDiagram}.  
The line-integrated forces do not match the originals for cases with caustics because there is no unique solution.
In theory, the power-diagram routine obtains the minimum deflection solution. 

\begin{figure}[htb]
\includegraphics[width=0.45\textwidth]{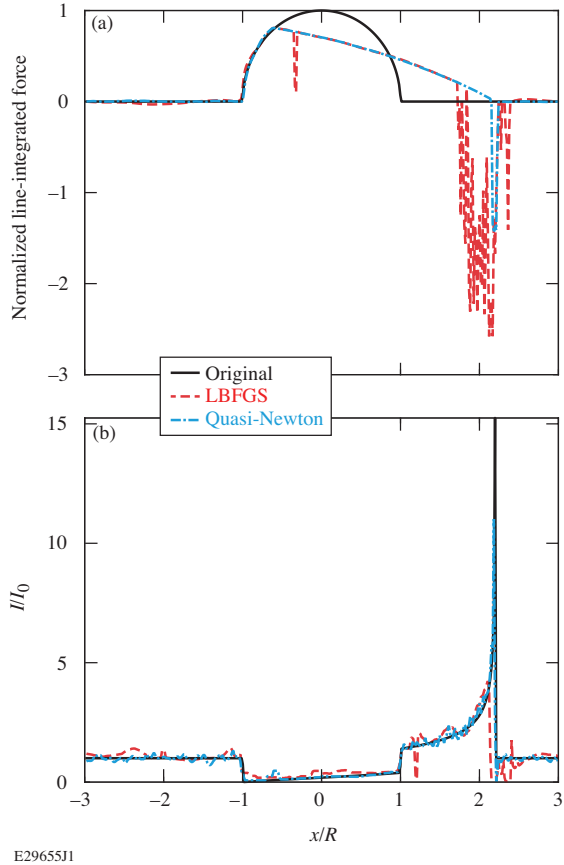}
\caption{\label{fig-PDfail} (a) Line-integrated fields normalized so that the maximum of the original is $1$, and (b) reconstructed proton radiographs from the power-diagram inversion routine for the top-hat profile with $\mu_{\max}=2$ and a bin width of $0.015R$ using two different minimization algorithms (LBFGS and quasi-Newton).}
\end{figure}

In two specific test cases with particularly sharp, single peaks the power-diagram routine failed to find an adequate solution.
The first case was the top-hat profile with $\mu_{\max} =2$ when we reduced the bin width from $0.025R$ to $0.015R$, which gives a sharper peak.
We did not carry out extensive tests on the effect of bin width, this run was carried out while choosing a bin width for this profile. 
Figure \ref{fig-PDfail} shows the reconstructions and line-integrated fields for this case.
Varying the number of sites used and the maximum number of iterations did not improve the solution.
The power-diagram routine has two options for the minimization algorithm: limited-memory Broyden-Fletcher-Goldfarb-Shanno (LBFGS), the default, and quasi-Newton gradient descent.    
Switching to quasi-Newton did improve the solution, which may appear to reproduce the radiograph in Fig.\ \ref{fig-PDfail}, but there is a sharp dip just beyond the peak that is not present in the original. 
Any feature distinct from those present in the original radiograph implies the presence of erroneous features in the line-integrated field.
In this case, a coarser binning led to an accurate solution, as seen in Fig.\ \ref{fig-PowerDiagram}. 
We also found that convolution with a Gaussian only a few bins wide led to an accurate solution, almost identical to that seen in Fig.\ \ref{fig-PowerDiagram}, when using the quasi-Newton algorithm.

\begin{figure}[htb]
\includegraphics[width=0.45\textwidth]{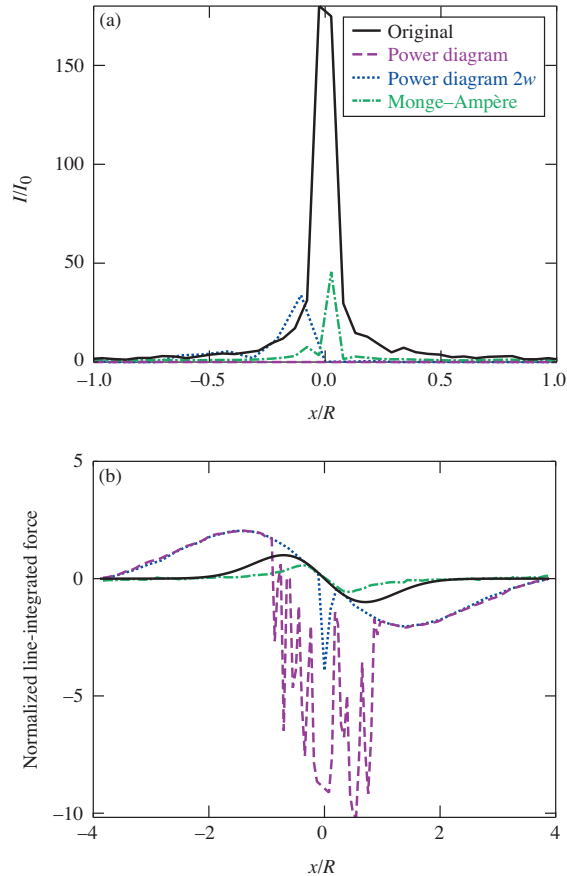}
\caption{\label{fig-GaussFail} Line-outs through the center of results for the spherical Gaussian potential with $\mu_{\max}=-0.5$ for (a) the proton radiographs, and (b) the line-integrated fields normalized so that the maximum of the original is $1$, $w$ refers to the bin width.}
\end{figure}

\begin{figure}[htb]
\includegraphics[width=0.9\textwidth]{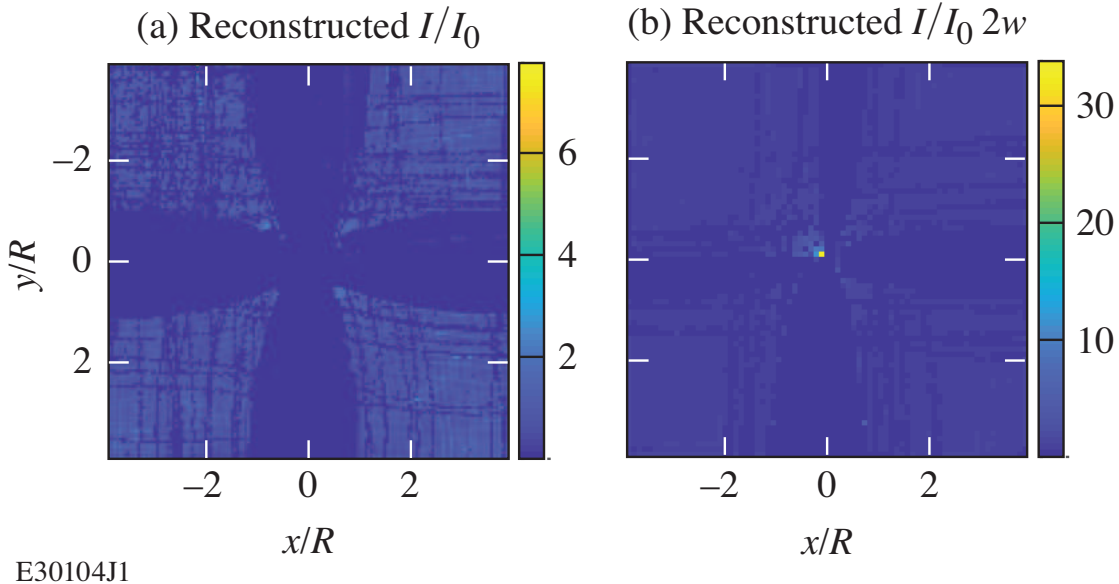}
\caption{\label{fig-GaussFail2D} Reconstructions from the power-diagram routine for the spherical Gaussian potential with $\mu_{\max}=-0.5$ for (a) a bin width $w$ of $0.052R$ (b) a bin width of $0.104R$.}
\end{figure}

The second test case for which the power-diagram routine failed was the spherical Gaussian potential with $\mu_{\max}=-0.5$, which produces a single peak since the two caustics are separated by less than the bin width.
The peak had a higher relative intensity than any of the other test cases. 
The power-diagram routine successfully inverted the remaining spherical Gaussian test cases.
Line-outs for this case are shown in Fig.\ \ref{fig-GaussFail}, using the quasi-Newton option.
We doubled the bin width and this gave a slight improvement, but the reconstruction is still inadequate.
The 2-D reconstruction, shown in Fig.\ \ref{fig-GaussFail2D}, demonstrates that the distortion introduced by the corners being forced to return to their original positions after converging to a solution is dominating the results.
In this case, the Monge-Amp\`{e}re routine came closer to the original, as seen in Fig.\ \ref{fig-GaussFail}, although the reconstructed peak is a factor of $3.4$ lower than the original.

We also carried out tests with added random noise and convolution with a Gaussian, which we will not reproduce here. 
The results for the spherical Gaussian potential shown in Fig.\ \ref{fig-failure}, which had a mean of only 10 particles per bin, demonstrate that the line-integrated forces are not affected by noise when an adequate reconstruction is obtained.
We found that adding a non-zero mean background level reduced the peaks in the line-integrated fields, but subtracting this mean level restored the original results.
We thought that convolution with a Gaussian might lead to the Monge-Amp\`{e}re routine converging to a more accurate result in the presence of caustics, but this did not occur for convolutions narrow enough to still allow key features in the radiographs to be distinguished.

During the course of this work, we found that the Monge-Amp\`{e}re routine can fail in the presence of empty bins in either the detected or source intensity, whereas the power-diagram can obtain a solution, although the results in empty regions were clearly inaccurate.

Finally, we will consider run time.
It is not possible to give a general consideration of run time because this is a strong function of the hardware used, particularly as the power-diagram routine is parallelized whereas the Monge-Amp\`{e}re routine is not.
The power-diagram routine running on a PC with 12 workers was always significantly slower than the Monge-Amp\`{e}re routine, up to $8000$ times slower in our tests, but only $1000$ times in cases without caustics.
The majority of the spherical Gaussian tests, which had $150\times150$ grids, took the power-diagram routine effectively a day to run, but the $\mu_{\max}=-0.5$ case took 32 hours, whereas the Monge-Amp\`{e}re routine rarely took more than a minute, taking 14 seconds for the $\mu_{\max}=-0.5$ case.
It is telling that the Monge-Amp\`{e}re routine has an inbuilt five minute wall time limit that we only discovered when reading through the code.
  
\section{Conclusions}
\label{sec-conclusions}

We have tested five direct inversion routines for proton radiography publicly available on GitHub.
The routines PRaLine\cite{Graziani} and PROBLEM\cite{PROBLEM} (both in Python) available at the time of writing did not run, and we did not attempt to fix them.
The routines invert\_shadowgraphy,\cite{powerDiagram} fast\_invert\_shadowgraphy,\cite{Fast} (both in Matlab) and PRNS\cite{PRNS} (Python) did run and we found them straightforward to use.
Both fast\_invert\_shadowgraphy\cite{Fast} and PRNS\cite{PRNS} use an implementation of the Sulman, Williams and Russell algorithm\cite{Sulman} by M. F. Kasim to solve the Monge-Amp\`{e}re equation.\cite{Monge}
PROBLEM also implements the same algorithm.
PRNS differs from the other routines in that it can determine a probability distribution for the source intensity starting from a given prior, assuming that there exists a unique solution for the line-integrated transverse Lorentz force.
The other routines require a specified source intensity assuming it to be uniform by default. 
The routines invert\_shadowgraphy and fast\_invert\_shadowgraphy have the option to use a source obtained by passing the data through a denoising algorithm, based on the assumption that large scale structures are inherent to the source.
We did not carry out tests with unspecified source intensities since the experiments that motivated this work used a D-$^3$He proton source that was found to be uniform in solid angle. 
We note, however, that the PRNS algorithm is predicated on the existence of a unique solution so do not expect it to work when caustics are present.

We used four different analytic radial force profiles to generate proton radiographs for a range of amplitudes, expressed as a dimensionless maximum force parameter $\mu$ [Eq.\ (\ref{eq-mu})].
Two of the cylindrical profiles -- linear and top-hat radial force profiles -- had discontinuities so they always produced caustics. 
They were chosen to be representative of the fields at ion fronts and material interfaces. 
Cylindrical and spherical Gaussian potentials that only produce caustics when $\mu_{\max} \ge 1.08$ or $\mu_{\max} \le -0.484$ were also used.
All of the radiographs generated, including those from the isothermal expansion model that we replaced with the linear profile in our tests, are included in the online supplementary material.

The Monge-Amp\`{e}re routines failed to reconstruct the test radiographs in most cases when caustics were present, even when the caustics were softened by convolution with a Gaussian.
The failure was obvious from the presence of erroneous features in the reconstructed radiographs, most notable of which were nonexistent, narrow, near-voids in proton intensity.
However, fast\_invert\_shadowgraphy did obtain a possible solution for the linear profile, ignoring an issue with the boundary conditions, albeit a nonphysical solution for the situation considered.

The power-diagram routine successfully inverted all but two of the test cases, which had particularly sharp, single peaks.
A top-hat force profile with a finer binning than our standard test cases was successfully inverted by use of either coarser binning or convolution with a Gaussian a few bins wide.
A spherical Gaussian potential that focused protons into a single peak could not be inverted.   
In the successful tests with caustics, the power-diagram routine gave a lower peak line-integrated field and a slower decay in the field than the original profiles, with the differences increasing with the magnitude of dimensionless force parameter $\mu$, as would be expected for a solution that minimizes proton deflection. 
We found that the power-diagram routine worked best with the quasi-Newton option (`algorithm', 'quasi-newton') for the minimization algorithm, which is not the default.
We also found that the power-diagram routine can still give an adequate solution with voids in the source or measured intensity, outside the region of the voids, which can break the Monge-Amp\`{e}re routines.

The boundary conditions in the routines we considered are predicated on the line-integrated forces and proton modulation going to zero.
The power-diagram routine can deal with forces parallel to the boundaries, except near the corners, which are forced back to their original positions after converging to a solution. 
The boundary conditions are particularly significant in the Monge-Amp\`{e}re routines we tested because they solve a diffusion-like equation, therefore errors can propagate over the entire grid.
Implementation of alternative boundary conditions allowing solutions with proton modulations at the boundary, such as 1-D cylindrical problems, would be useful. 
It would also be desirable to eliminate the distortion at the corners in the power-diagram routine, even if that meant obtaining a solution over a smaller region of the detector.
The routines also require that all deflected protons be detected, which means that the force perpendicular to the boundaries must go to zero and that the forces even well within the boundaries cannot be too large. 
We have found, from a limited number of cases, that missing protons can lead to erroneous fields near the boundaries. 

The power-diagram routine can be very slow; the spherical Gaussian tests with $150\times150$ grids required essentially a day to invert using a PC doing nothing else. 
An implementation in a compiled rather than an interpreted language running on a machine with at least hundreds of cores would be desirable, which is possible with the routine provided.
For proton radiographs without caustics, the Monge-Amp\`{e}re routines have a considerable advantage in speed despite not running in parallel, up to $1000$ times in our tests, never taking more than a few minutes, which makes them the ideal analysis tool for proton radiographs without caustics.

It would be of interest to explore algorithms capable of generating a range of possible solutions subject to a variety of physical constraints in the presence of caustics.
For example, in addition to minimum deflection, minimum field energy would be a physically relevant constraint.
Machine-learning algorithms may be able to achieve these goals if adequately trained with suitable physical models of the cases to be analyzed.\cite{machineLearning}

Finally, it should not be forgotten that direct inversion can be applied to radiography with any charged particle and to shadowgraphy, where photons are deflected by refractive index gradients causing intensity modulations.
All of these diagnostic techniques could be referred to generically as deflectometry.

\section{Data Availability}

The data that support the findings of this study are available in the online supplementary material.

\section{Supplementary Material}

All of the proton radiographs referred to in this paper are available as HDF files using pradformat\cite{pradformat} in the online supplementary material. 

\appendix

\section{Approximate solution to the cylindrical Murakami-Basko equation}

Murakami and Basko\cite{MB} derive a self-similar equation for a 1-D collisionless plasma expansion with cold ion ions and Maxwellian electrons without making the quasi-neutral approximation. 
A self-similar solution is only possible in this case when the self-similar scale length and the electron Debye length maintain a constant ratio, giving only one independent length scale to the problem, which requires a specific adiabatic index for the electrons.
For a cylindrical expansion the electrons must be isothermal, which is relevant to a laser heated cylindrical plasma.
The cylindrical and spherical cases cannot be solved analytically. 
Murakami and Basko use the 1-D planar solution as an approximation to these cases.
We have found an improved approximation to the cylindrical case, \eq{eq-Eiso}, assuming $\Lambda \gg 1$ and $Zm_{\rm e} \ll m_{\rm i}$, where $Z$ is ion charge number, $m_{\rm e}$ is electron mass, and $m_{\rm i}$ is ion mass.
The assumption $Zm_{\rm e} \ll m_{\rm i}$ allows a term that results from electron dynamics to be neglected.
The electron and ion density profiles in this limit are Gaussian with $1/$e radii of $R/\sqrt{\ln\Lambda}$ (which Murakami and Basko refer to as $R$) with the ion density having a sharp cutoff at $R$, which was the basis for our approximation.
The physical values of $E_{\max}$ and $\Lambda$ are not required for generating proton radiographs, where $E_{\max}$ is replaced by $\mu_{\max}$ and $\ln\Lambda$ is a free dimensionless parameter describing the field profile in the electron sheath.
However, for the sake of providing a complete physical picture, we have
\begin{eqnarray}
\label{eq-Lambda}
\Lambda & = & \frac{R^2/(4\lambda_{\rm D}^2\ln\Lambda)}{\ln[R^2/(4\lambda_{\rm D}^2\ln\Lambda)]}, \\
\label{eq-Emax}
E_{\max} & = & \frac{2kT}{qR}\ln\Lambda,
\end{eqnarray}   
where $k$ is Boltzmann's constant, $T$ is electron temperature, and $\lambda_{\rm D}$ is electron Debye length on-axis. 
An explicit solution for the key parameter $\ln\Lambda$ cannot be obtained.
Note that $R/\lambda_{\rm D}$ and hence $\Lambda$ have constant values while $R$ increases in time according to
\begin{equation}
\frac{dR}{dt} = 2c_{\rm s} \sqrt{(\ln\Lambda) \ln\frac{R}{R_0}},
\end{equation}
where $c_{\rm s}$ is the isothermal ion sound speed, which cannot be solved explicitly for $R$. 

\acknowledgements

The authors thank A. F. A. Bott for pointing them to the fast\_invert\_shadowgraphy routine and S. Feister for providing and assisting them with pradformat.
The information, data, or work presented herein was funded in part by the Advanced Research Projects Agency-Energy (ARPA-E), U.S. Department of Energy, under Award Number DE-AR0000568, by the Department of Energy National Nuclear Security Administration under Award Number DE-NA0003856, the University of Rochester, and the New York State Energy Research and Development Authority. This report was prepared as an account of work sponsored by an agency of the U.S. Government. Neither the U.S. Government nor any agency thereof, nor any of their employees, makes any warranty, express or implied, or assumes any legal liability or responsibility for the accuracy, completeness, or usefulness of any information, apparatus, product, or process disclosed, or represents that its use would not infringe privately owned rights. Reference herein to any specific commercial product, process, or service by trade name, trademark, manufacturer, or otherwise does not necessarily constitute or imply its endorsement, recommendation, or favoring by the U.S. Government or any agency thereof. The views and opinions of authors expressed herein do not necessarily state or reflect those of the U.S. Government or any agency thereof.


\begin{thebibliography}{99}

\bibitem{Graziani} C. Graziani, P. Tzerferacos, D. Q. Lamb, and C.-K. Li, Rev. Sci. Instr. {\bf 88}, 123507 (2017); PRaLine (Proton Radiography Linear reconstruction), accessed 30 August 2021, github.com/flash-center/PRaLine

\bibitem{powerDiagram} Muhammad Firmansyah Kasim, Luke Ceurvorst, Naren Ratan, James Sadler, Nicholas Chen, Alexander Savert, Raoul Trines, Robert Bingham, Philip N. Burrows, Malte C. Kaluza, and Peter Norreys, Phys. Rev. E {\bf 95}, 023306 (2017); Invert Shadowgraphy and Proton Radiography, accessed 8 July 2021, github.com/mfkasim1/invert-shadowgraphy.

\bibitem{Fast} M. F. Kasim, Invert Shadowgraphy and Proton Radiography, accessed 8 July 2021, github.com/mfkasim1/invert-shadowgraphy/tree/fast-inverse; M. Sulman, J. F. Williams, and Robert D. Russell, Appl. Numer. Math. {\bf 61}, 298 (2011).

\bibitem{PROBLEM} A. F. A. Bott, C. Graziani, P. Tzeferacos, T. G. White, D. Q. Lamb, G. Gregori, and A. A. Schekochihin, J. Plasma Physics {\bf 83}, 905830614 (2017); PROBLEM Solver (PROton-imaged B-field nonLinear Extraction Module), accessed 12 July 2021, github.com/flash-center/PROBLEM.

\bibitem{PRNS} M. F. Kasim, A. F. A. Bott, P. Tzeferacos, D. Q. Lamb, G. Gregori, and S. M. Vinko, Phys. Rev. E {\bf 100}, 033208 (2019); PRNS (Proton Radiography with No Source), accessed 12 July 2021, github.com/OxfordHED/proton-radiography-no-source.

\bibitem{machineLearning} Nicholas F. Y. Chen, Muhammad Firmansyah Kasim, Luke Ceurvorst, Naren Ratan, James Sadler, Matthew C. Levy, Raoul Trines, Robert Bingham, and Peter Norreys, Phys. Rev. E {\bf 95}, 043305 (2017).

\bibitem{Monge} Gaspard Monge, Mémoire sur la théorie des déblais et des remblais, Histoire de l’Académie royale des sciences avec les mémoires de mathématique et de physique tirés des registres de cette Académie (1781), 666-705. Commentary and extracts in French, accessed 1 November 2021: http://images.math.cnrs.fr/Gaspard-Monge,1094.html?lang=fr

\bibitem{Slutz} S. A. Slutz, M. C. Herrmann, R. A. Vesey, A. B. Sefkow, D. B. Sinars, D. C. Rovang, K. J. Peterson, and M. E. Cuneo, Phys. Plasmas {\bf 17}, 056303 (2010).

\bibitem{miniMagLIF} D. H. Barnak, J. R. Davies, R. Betti, M. J. Bonino, E. M. Campbell, V. Y. Glebov, D. R. Harding, J. P. Knauer, S. P. Regan, A. B. Sefkow, A. J. Harvey-Thompson, K. J. Peterson, D. B. Sinars, S. A. Slutz, M. R. Weis, and P.-Y. Chang, Phys. Plasmas {\bf 24}, 056310 (2017); J. R. Davies, D. H. Barnak, R. Betti, E. M. Campbell, P.-Y. Chang, A. B. Sefkow, K. J. Peterson, D. B. Sinars, and M. R. Weis, Phys. Plasmas {\bf 24}, 062701 (2017); E. C. Hansen, J. R. Davies, D. H. Barnak, J. Peebles, A. B. Sefkow, P.-Y. Chang, R. Betti, V. Yu. Glebov, J. P. Knauer, E. M. Campbell, and S. P. Regan, Plasma Phys. Control. Fusion {\bf 60}, 054014 (2018); J. R. Davies, R. Bahr, D. H. Barnak, R. Betti, M. J. Bonino, E. M. Campbell, E. C. Hansen, D. R. Harding, J. L. Peebles, A. B. Sefkow, W. Seka, P.-Y. Chang, M. Geissel, and A. J. Harvey-Thompson, Phys. Plasmas {\bf 25}, 062704 (2018); E. C. Hansen, D. H. Barnak, P.-Y. Chang, R. Betti, E. M. Campbell, J. R. Davies, J. P. Knauer, J. L. Peebles, S. P. Regan, and A. B. Sefkow, Phys. Plasmas {\bf 27}, 062703 (2020); D. H. Barnak, M. J. Bonino, P.-Y. Chang, J. R. Davies, E. C. Hansen, D. R. Harding, J. L. Peebles, and R. Betti, Phys. Plasmas {\bf 27}, 112709 (2020).

\bibitem{PlasmaPy} PlasmaPy Community, E. Everson, D. Stańczak, N. A. Murphy, P. M. Kozlowski, R. Malhotra, S. J. Langendorf, A. J. Leonard, D. Stansby, C. C. Haggerty, S. J. Mumford, J. P. Beckers, M. S. Bedmutha, J. Bergeron, L. Bessi, K. Bryant, S. Carroll, S. Chambers, A. Chattopadhyay, A. Choubey, J. Deal, D. Diaz, R. Díaz Pérez, L. Einhorn, T. Fan, S. I. Farid, G. Goudeau, S. Guidoni, R. S. Hansen, P. Heuer, J. Hillairet, P. Z. How, Y.-M. Huang, N. Humphrey, M. Isupova, J. Kent, S. Kulshrestha, P. Kuszaj, P. L. Lim, A. Magarde, J. V. Martinelli, J. Munn, T. Parashar, N. Patel, J. Polak, A. Rao, R. Raj, V. Rajashekar, A. Savcheva, C. Shen, D. N. Sherpa, F. Silva, A. Singh, A. Singh, B. Sipőcz, A. Tavant, T. Varnish, A. Vo, S. Xu, C. Zhang, T. Du, R. Qudsi, S. Richardson, C. Skinner, D. Modi, D. Drozdov, and K. Montes, PlasmaPy (Version 0.6.0), Zenodo, Accessed 14 March 2021, http://doi.org/10.5281/zenodo.4602818.

\bibitem{MB} M. Murakami and M. M. Basko, Phys. Plasmas {\bf 13}, 012105 (2006).

\bibitem{Us} J. R. Davies, R. Betti, P.-Y. Chang, and G. Fiksel, Phys. Plasmas {\bf 22}, 112703 (2015).

\bibitem{Sulman} Mohamed M. Sulman, J. F. Williams, and Robert D. Russell, Applied Numerical Mathematics {\bf 61} 298 (2011).

\bibitem{pradformat} S. Feister, Pradformat (Proton Radiography File Format Tools), accessed 12 July 2021, github.com/phyzicist/pradformat.

\end{thebibliography}
\end{document}